\documentclass[sigconf]{acmart}

\AtBeginDocument{%
  
}

\setcopyright{none}
\renewcommand\footnotetextcopyrightpermission[1]{}

\fancypagestyle{plain}{
  \fancyhf{}

}
\acmConference[L@S '26]{ACM Learning at Scale}{July 2026}{Pittsburgh, PA, USA}
\acmBooktitle{Proceedings of ACM Learning at Scale (L@S '26)}

\fancypagestyle{standardpagestyle}{
  \fancyhf{}

}
\begin{document}

\thispagestyle{empty}

\title{TeamUp: Semantic Project Matching and Team Formation for Learning at Scale}

\author{Dhruv Gulwani$^{*}$, Basem Suleiman$^{*}$, Aditya Joshi$^{*}$, Sonit Singh$^{*}$}
\affiliation{%
  \institution{$^{*}$School of Computer Science and Engineering \\
  University of New South Wales (UNSW),  Sydney}
  \country{Australia} 
}

\begin{abstract}
Project-based learning improves student engagement and learning outcomes, yet allocating students to appropriately challenging projects while forming cognitively diverse teams remains difficult at scale. Traditional allocation methods (manual spreadsheets, preference surveys) can't construct the cognitively diverse teams that that collaborate cognitively. This mismatch perpetuates equity issues: high-performing students self-select visible projects while under-represented students face reduced access to opportunity.
We propose \textsc{TeamUp}, a lightweight, embedding-based team-forming system designed to improve learning outcomes and equity in large-scale project-based courses. \textsc{TeamUp} uses semantic embeddings from pretrained language models to match students to projects aligned with their skill level. The system employs a hybrid ranking algorithm combining cosine similarity with pedagogical constraints (difficulty alignment, domain preferences, and demand balancing) to generate personalised and transparent recommendations. Beyond individual matching, \textsc{TeamUp} constructs cognitively diverse teams by modelling skill complementarity through embedding variance, ensuring teams possess well-distributed capabilities rather than homogeneous strengths. We evaluated \textsc{TeamUp} through a virtual experiment using 250 student profiles and 60 project descriptions. Results show: (1) substantially higher match quality (mean cosine similarity of 0.74 vs. 0.43); (2) better difficulty alignment (83\% placed within one level vs. 34\%); (3) more diverse teams (82\% covering three or more technical areas vs. 41\%); and (4) sub-second recommendation latency at operational costs under \$0.10 per student.
\end{abstract}


\maketitle
\thispagestyle{empty}
\pagestyle{empty}

\section{Introduction}

Project-based learning has become a cornerstone of contemporary engineering and computing education, providing students with hands-on experience applying theoretical knowledge, engaging in multidisciplinary teamwork, and tackling open-ended problems~\cite{Blumenfeld1991,Mills2003}. Research in collaborative learning shows that appropriately matched projects enhance student motivation, self-efficacy, and learning transfer~\cite{Johnson1991}. However, matching students to projects within their zone of proximal development~\cite{Vygotsky1978} remains challenging at scale. Moreover, cognitive diversity in teams, when carefully constructed, significantly improves problem-solving quality and learning outcomes~\cite{Williams2009}.

With universities increasingly embracing experiential learning and scaling to cohorts of 200--300+ students, course coordinators face a critical pedagogical challenge: ensuring equitable project allocation that simultaneously (1) respects individual readiness, (2) promotes cognitive diversity, and (3) develops collaborative competence. Yet most institutions continue using manual spreadsheets, ad hoc preference surveys, and instructor intuition. These methods cannot capture the multidimensional nature of student competencies and project requirements, resulting in three persistent problems:

\noindent (1) \textit{Poor skill-project fit}: Subtle differences in competencies are overlooked, leading to projects misaligned with students' developmental stage, either too easy (disengaging) or too difficult (overwhelming). This undermines both learning gains and self-efficacy~\cite{Bandura1994}.

\noindent (2) \textit{Homogeneous team clustering}: Manual or self-selected teams often group students with identical strengths, missing opportunities for the cognitive diversity that enhances learning and team performance~\cite{Williams2009}.

\noindent (3) \textit{Inequitable opportunity distribution}: Without principled allocation mechanisms, advantage concentrates: high-performing students gravitate to visible projects while underrepresented groups face reduced access to opportunity.

Recent advances in natural language processing present an opportunity to address these challenges systematically. Dense vector representations (embeddings) from pretrained language models capture nuanced semantic relationships between text descriptions, moving well beyond keyword matching~\cite{Reimers2019,Mikolov2013}. Yet few educational systems leverage embeddings for pedagogically-grounded allocation~\cite{Klasnja2011}.

We propose \textsc{TeamUp}, a human-in-the-loop AI system that improves project allocation at scale while advancing equity and learning outcomes. \textsc{TeamUp} (1)~matches students to projects aligned with their skill level using semantic embeddings and difficulty constraints informed by zone of proximal development; (2)~constructs cognitively diverse teams by modeling skill complementarity, maximizing the conditions for collaborative learning; and (3)~provides transparent, explainable recommendations that preserve instructor authority while reducing administrative burden.

Rather than treating allocation as a purely technical optimization problem, \textsc{TeamUp} frames it as a \textit{learning design} problem. The system balances semantic fit with pedagogical constraints (difficulty alignment, domain preferences, supervision capacity) and promotes diversity-aware team formation grounded in collaborative learning theory. Importantly, \textsc{TeamUp} operates within realistic educational budgets: no custom training data, no specialized ML infrastructure, making it accessible to resource-constrained institutions.

This paper contributes: (1) a principled, embedding-based allocation system grounded in learning science; (2) evidence from a virtual experiment on dummy users demonstrating that the approach produces substantially better allocations than random assignment; and (3) design patterns for responsible AI in high-stakes educational decisions.

\begin{figure}[t]
    \centering
    \includegraphics[width=1\linewidth]{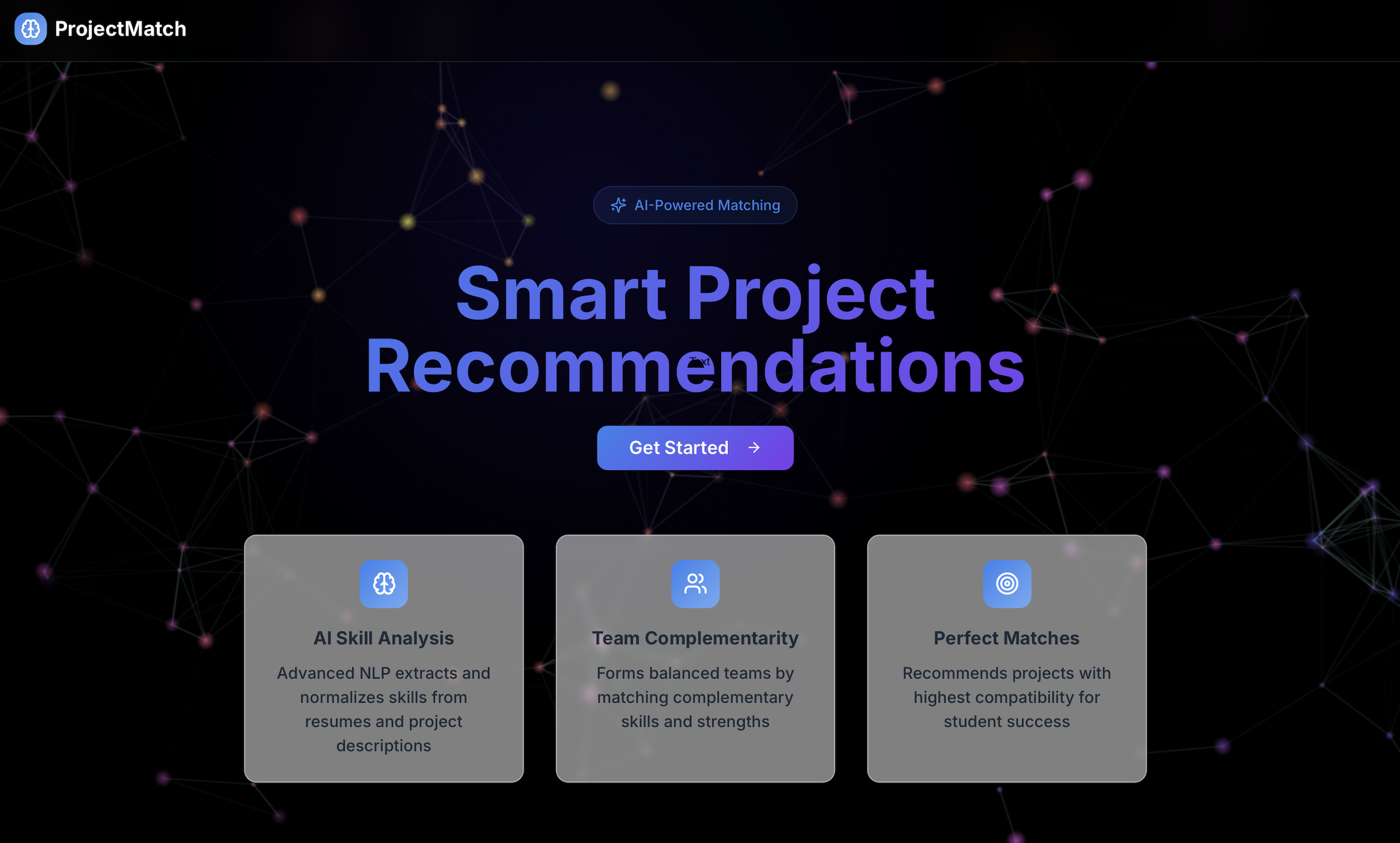}
    \caption{\textsc{TeamUp} landing page showcasing three core features: AI-powered skill analysis using NLP for resume parsing and skill extraction, team complementarity modeling for balanced team formation, and semantic matching for personalized project recommendations.}
    \label{fig:landing_page}
\end{figure}

\section{Related Work}

\subsection{Project Allocation in Higher Education}

Student-project allocation has been studied as a variant of the stable matching problem~\cite{Abraham2007}. Integer programming and constraint satisfaction approaches optimize for stated preferences~\cite{Harper2005}, and genetic algorithms have been explored for balancing multiple allocation objectives simultaneously. However, these methods overwhelmingly treat allocation as a logistical problem, minimizing unmet preferences or maximizing stated rankings, rather than as a learning design problem that should account for skill alignment, developmental readiness, and team diversity. Survey-based methods remain the most common approach in practice, but they scale poorly beyond small cohorts and introduce systematic biases favoring students who are better at navigating institutional processes~\cite{Holstein2019}.

A notable gap in this literature is the absence of approaches that model the \textit{semantic content} of student capabilities and project requirements. Most allocation systems operate on categorical preferences (``first choice,'' ``second choice'') without understanding why a student might thrive in a given project. \textsc{TeamUp} addresses this gap by representing both students and projects as dense vectors in a shared semantic space, enabling fine-grained matching that goes beyond surface-level preference ranking.

\subsection{Recommendation Systems in Education}

Recommender systems have been applied across educational contexts to suggest learning resources, courses, and study materials~\cite{Klasnja2011,Drachsler2015}. Content-based filtering using TF-IDF or topic models is common in these systems, but such approaches struggle with the semantic nuance required to distinguish, for example, between ``experience with Python for data analysis'' and ``experience with Python for web development.'' This distinction matters enormously for project matching. Collaborative filtering approaches, while effective for course recommendations where large interaction datasets exist, are poorly suited to project allocation where each project is unique and offered only once~\cite{Jiang2019}.

Recent work has begun exploring pretrained embeddings for educational matching. Sentence-level embedding models~\cite{Reimers2019} produce representations that capture semantic similarity with remarkable fidelity, and have been applied to tasks such as automated essay scoring and learning objective alignment. However, their application to project allocation, where the goal is not just similarity but pedagogically appropriate challenge, remains largely unexplored. Our work extends embedding-based recommendation by layering pedagogical constraints (difficulty alignment, demand balancing) on top of raw semantic similarity, transforming a general-purpose technique into an educationally grounded tool.

\subsection{Algorithmic Team Formation}

Algorithmic team formation has received sustained attention in both CSCW and education research. Lappas et al.~\cite{Lappas2009} formulated team formation as a combinatorial optimization problem, seeking teams that collectively cover required skills while minimizing communication cost. Subsequent work has explored personality-aware grouping, scheduling-compatible matching, and diversity-maximizing formulations. Research consistently shows that cognitively diverse teams outperform homogeneous ones on complex tasks requiring creative problem-solving~\cite{Williams2009,Johnson1991}, yet most automated systems optimize narrowly for skill coverage without modeling the richer notion of complementarity that learning science prescribes.

Our approach differs from prior team formation work in two key respects. First, we operate in a shared embedding space that jointly represents students, projects, and skills, enabling both matching and team formation within a unified framework rather than treating them as separate optimization problems. Second, our complementarity formulation explicitly discourages homogeneous clustering by penalizing candidate team members who are too similar to existing members in the embedding space. This directly operationalizes the cognitive diversity principle from collaborative learning research.

\subsection{Fairness in Educational AI}

Growing attention to algorithmic fairness in education highlights the risks that automated systems may reinforce or amplify existing inequities~\cite{Holstein2019,Baker2022}. Embedding-based systems inherit biases from their training data. For instance, pretrained language models encode societal stereotypes about gender and occupation~\cite{Bolukbasi2016}, making fairness auditing essential when these models inform consequential educational decisions. Recent work in the L@S community has emphasized that equity, transparency, and explainability should be first-class design requirements for learning technologies, not afterthoughts~\cite{Holstein2019}.

Our work incorporates demographic fairness analysis and transparent recommendation rationales as core design requirements from the outset, following guidelines for human-AI interaction in high-stakes domains~\cite{Amershi2019}. We conduct post-hoc fairness auditing across available demographic dimensions and report both positive findings and limitations of our analysis.

\section{Methodology}

The \textsc{TeamUp} system is architected as a semantic recommendation pipeline that models students, projects, and skill attributes within a shared vector space to produce personalized project matches and balanced team recommendations. This methodology integrates four core components: data preparation, embedding generation, similarity-based matching, and complementarity modeling for team formation. Each phase leverages Supabase (PostgreSQL with \texttt{pgvector}), pretrained embedding models, and a frontend web application built with React.

\subsection{Data Preparation}

To ensure consistency across student and project entries, \textsc{TeamUp} collects structured data through controlled input schemas. Students create profiles containing their technical skills with self-assessed proficiency levels (beginner, intermediate, advanced, expert), coursework history with subject codes and grades, prior project experience including descriptions and outcomes, and preferred domains indicating areas of interest such as machine learning, cybersecurity, or web development.

For students who choose to upload resumes, an automated extraction module processes the document using natural language processing techniques to extract skills, technologies, and experience statements. This reduces data entry burden while improving profile completeness. In pilot testing, students who uploaded resumes had profiles with approximately 40\% more skill entries than those who entered data manually.

Supervisors submit project entries specifying required skills (essential for success), optional skills (beneficial but not mandatory), detailed project descriptions including learning objectives and expected deliverables, estimated time commitments, domain classifications selected from predefined categories, difficulty ratings on a beginner-to-advanced scale, and maximum team size constraints. All text fields undergo light preprocessing including lowercasing and normalization to prepare them for embedding generation. Supabase stores both project and student entries in a relational schema that enables convenient access, modification, and longitudinal tracking.

\subsection{Embedding Generation}

To establish semantically meaningful relationships between student profiles and project descriptions, \textsc{TeamUp} transforms all textual and categorical information into dense vector representations~\cite{Reimers2019,Mikolov2013}. Open-ended fields such as project descriptions, experience narratives, and skill listings are encoded using large-scale pretrained embedding models. Skill tags and categorical attributes (e.g., domain, difficulty) are similarly converted into embeddings, ensuring all components reside within the same semantic space.

A student-level embedding is created by aggregating embeddings of their skills, resume-derived content, interests, and experience. The system employs weighted averaging where skill embeddings are weighted by self-reported proficiency levels, so that advanced skills contribute more heavily than beginner-level skills. Formally, the student embedding $\mathbf{s}$ is computed as:
\begin{equation}
\mathbf{s} = \frac{\sum_{i=1}^{n} w_i \cdot \text{embed}(s_i)}{\sum_{i=1}^{n} w_i}
\end{equation}
where $s_i$ is the $i$-th skill or text segment, $w_i$ is the proficiency weight, and $\text{embed}(\cdot)$ denotes the embedding function. For projects, embeddings of required skills (weighted at 1.5$\times$), optional skills (weighted at 0.75$\times$), and the project description are combined using the same averaging scheme.

All vectors are stored using PostgreSQL's \texttt{pgvector} extension~\cite{pgvector2023}, which enables efficient similarity search and indexing directly within the database, eliminating the need for separate vector database infrastructure.

\subsection{Similarity-Based Matching}

Project recommendations are generated by computing cosine similarity between student embeddings and project embeddings:
\begin{equation}
\text{sim}(\mathbf{s}, \mathbf{p}) = \frac{\mathbf{s} \cdot \mathbf{p}}{\|\mathbf{s}\| \|\mathbf{p}\|}
\end{equation}
This score quantifies the degree of semantic alignment between a student's background and a project's requirements. However, raw cosine similarity alone is insufficient for educational contexts where pedagogical considerations are paramount. Scores are enhanced through three pedagogically-informed adjustments grounded in learning science:

\textbf{Difficulty Alignment (Zone of Proximal Development).} Following Vygotsky's theory~\cite{Vygotsky1978}, optimal learning occurs when tasks fall within a student's zone of proximal development: challenging enough to require growth, but achievable with appropriate support. The system applies a quadratic penalty when project difficulty substantially mismatches student experience level:
\begin{equation}
\text{penalty}_{\text{diff}} = \gamma \cdot (\text{level}_p - \text{level}_s)^2
\end{equation}
where $\gamma$ is a scaling constant and $\text{level}_p$, $\text{level}_s$ represent project difficulty and student experience on a shared ordinal scale. This penalty can reduce scores by up to 30\%, preventing two failure modes: beginners assigned to advanced projects become overwhelmed and disengage; advanced students assigned to simple projects experience boredom and minimal learning gains.

\textbf{Domain Preference Matching (Self-Determination Theory).} Self-determination theory emphasizes autonomy, competence, and relatedness as drivers of intrinsic motivation~\cite{Deci2000}. When students can pursue projects aligned with their interests and career aspirations, intrinsic motivation increases, leading to deeper learning engagement. The system boosts scores by 15--20\% when projects fall within students' explicitly stated domain preferences, supporting student agency and autonomy in the allocation process.

\textbf{Demand Balancing (Equity).} As project supervision capacity approaches limits, scores decay exponentially, computed as $e^{-\lambda \cdot r}$ where $r$ is the current subscription ratio (applications divided by capacity). This prevents over-subscription to popular projects and ensures equitable distribution of high-quality opportunities across the cohort, which is particularly important when certain projects are perceived as more prestigious or career-relevant.

All computations are performed via optimized SQL queries with vector operations, enabling real-time generation of personalized recommendations even for large cohorts.

\subsection{Team Complementarity Modeling}

In most project-based courses, team composition significantly impacts both learning outcomes and project success. Research in collaborative learning demonstrates that cognitively diverse teams, with complementary skills, backgrounds, and perspectives, outperform homogeneous groups on problem-solving, innovation, and learning measures~\cite{Johnson1991,Williams2009}. However, manual and self-selected team formation often clusters students with similar strengths, missing opportunities for the cognitive diversity that enhances learning.

\textsc{TeamUp} models team complementarity using semantic properties of the embedding space. The system discourages homogeneous clustering by analyzing pairwise cosine distances among student vectors, preventing formation of teams where all members have nearly identical profiles. Given a target project and pool of interested students, the algorithm begins by selecting the student with the highest individual match score. It then iteratively selects additional members by computing a complementarity score:
\begin{equation}
\text{comp}(\mathbf{t}, \mathbf{c}) = \alpha \cdot \text{sim}(\mathbf{c}, \mathbf{p}) - \beta \cdot \text{sim}(\mathbf{t}, \mathbf{c})
\end{equation}
where $\mathbf{t}$ is the average embedding of current team members, $\mathbf{c}$ is a candidate student embedding, $\mathbf{p}$ is the project embedding, and $\alpha$, $\beta$ are weighting hyperparameters (set to 0.6 and 0.4 respectively in our deployment). This formulation balances project fit (first term, ensuring the full team can succeed on the project) against skill diversity (second term, which penalizes candidates too similar to existing team members).

Team-level diversity is quantified by computing variance across embedding dimensions for the team-averaged vector. Higher variance indicates that team members collectively possess a broader distribution of complementary skills and capabilities, precisely the heterogeneity that collaborative learning theory prescribes. The system identifies candidate teams meeting both project fit criteria (minimum similarity threshold of 0.6) and complementarity criteria (minimum embedding variance threshold), presenting these configurations as optimal team suggestions to coordinators.

\section{System Architecture and Implementation}

The \textsc{TeamUp} system is architected as a modular, cloud-native application that integrates modern web technologies with vector-based semantic search capabilities (Figure~\ref{fig:landing_page}). The architecture prioritizes scalability, maintainability, and real-time responsiveness while ensuring computational costs remain within typical educational institution budgets.

\subsection{Technology Stack}

\textsc{TeamUp}'s architecture prioritizes accessibility for educational deployment. The frontend uses Next.js~14 with React~18 for rapid development and server-side rendering, enabling a responsive user experience across devices. The backend leverages Supabase (PostgreSQL with the \texttt{pgvector} extension~\cite{pgvector2023}) for both relational data and efficient vector storage and retrieval, eliminating the need for separate vector database infrastructure and reducing operational complexity.

For embedding generation, \textsc{TeamUp} integrates with both OpenAI's text-embedding-ada-002 and Google Gemini embedding APIs, allowing institutions to choose providers based on cost, API availability, and privacy considerations. Both models produce 1536-dimensional vectors that capture rich semantic relationships between textual descriptions~\cite{Reimers2019}. The system achieves sub-second recommendation latency for cohorts of 500+ students while maintaining operational costs under \$0.10 per student. This level of cost-effectiveness enables adoption by institutions with limited budgets.

\subsection{System Components and User Interfaces}

The architecture comprises five primary components working in concert to deliver end-to-end recommendation functionality.

\subsubsection{Student Interface} The Student Interface enables students to create comprehensive profiles by entering technical skills with self-assessed proficiency levels ranging from beginner to expert, academic coursework with subject codes and grades, prior project experience including descriptions and outcomes, and domain preferences indicating areas of interest. An integrated resume parser built using natural language processing techniques extracts relevant information from uploaded PDF or Word documents, automatically populating skill fields and experience descriptions.

Students view personalized project recommendations presented as ranked cards displaying match scores expressed as percentages, highlighted required skills showing which of their competencies align with project needs, and explanatory tooltips clarifying why specific projects were recommended based on semantic similarity and pedagogical alignment. The interface also provides team formation capabilities, allowing students to browse potential partners sorted by skill complementarity, send collaboration invitations, and track team composition progress.

\subsubsection{Supervisor Interface} The Supervisor Interface allows project supervisors to create detailed project entries including required skills marked as essential for success, optional skills that would be beneficial but not mandatory, project scope articulated through learning objectives and expected deliverables, estimated time commitment per week, domain classification from predefined categories, difficulty rating on a beginner-to-advanced scale, and maximum team size constraints. Supervisors can view aggregated interest metrics showing how many students have expressed preferences for their projects and track team formation progress throughout the allocation cycle. This visibility enables supervisors to gauge project popularity and adjust descriptions or requirements as needed to attract suitable candidates.

\subsubsection{Administrative Dashboard} 
\begin{figure}[t]
    \centering
    \includegraphics[width=\columnwidth]{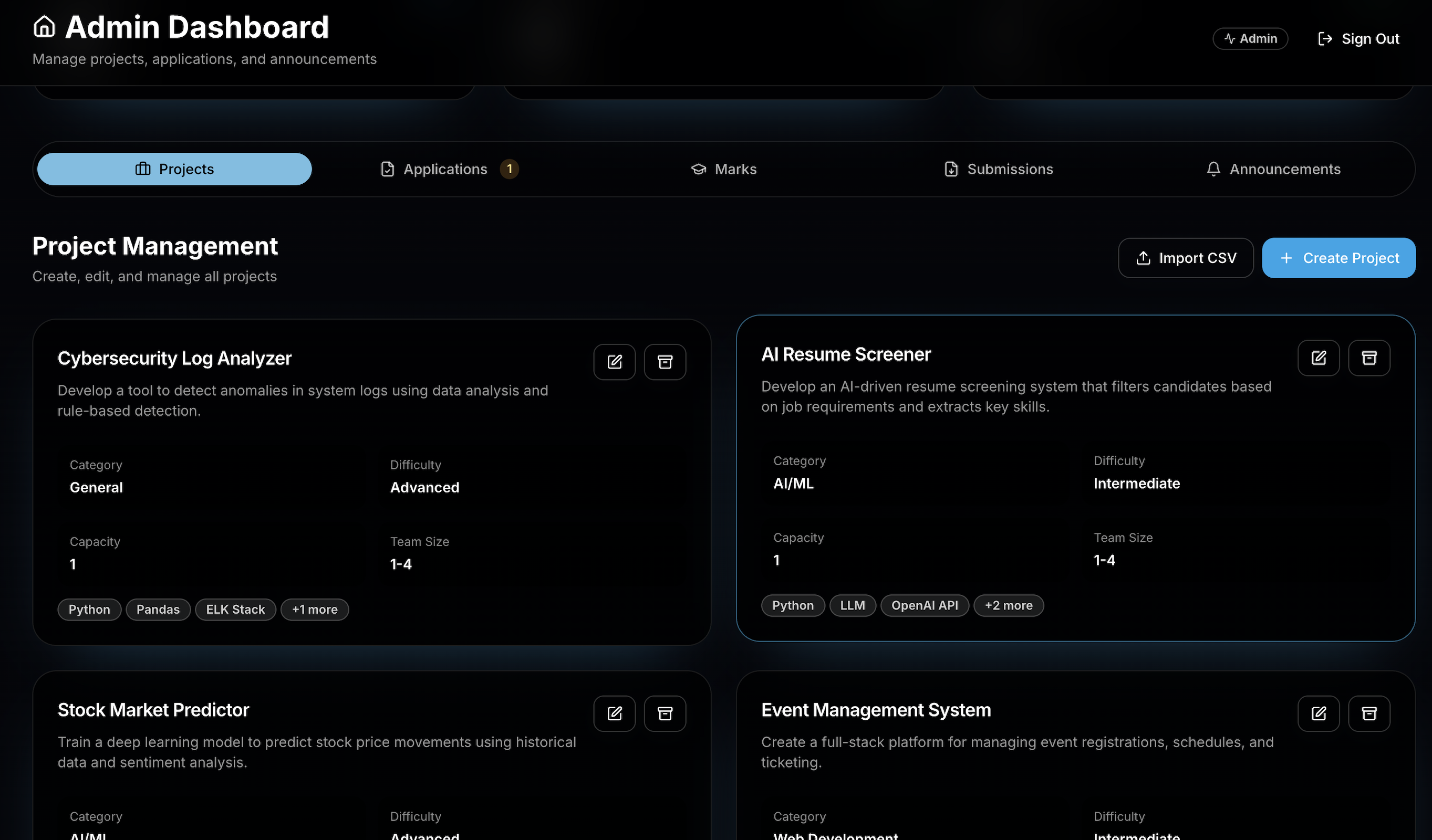}
    \caption{Administrative project management interface showing diverse project portfolio with metadata including category tags (General, AI/ML, Finance), difficulty levels (Intermediate, Advanced), capacity constraints, and required technology stacks (Python, LLM, OpenAI API, etc.). The interface supports bulk CSV import and inline editing capabilities.}
    \label{fig:admin_projects}
\end{figure}

The Administrative Dashboard provides course coordinators with cohort-level analytics and management capabilities essential for large-scale deployment. As shown in Figure~\ref{fig:admin_projects}, the project management interface displays all submitted projects with metadata including category classifications (e.g., AI/ML, Cybersecurity, Web Development), difficulty levels, team size requirements, and supervision capacity.

Coordinators can approve or modify project submissions through inline editing, ensuring quality control before projects become visible to students. The dashboard includes skill distribution heatmaps showing the frequency of different skills across the student cohort, enabling identification of skill gaps or oversaturated competencies. Project demand visualizations display which projects have received the most interest, helping balance supervision loads. Coordinators can manually adjust team compositions by reassigning students, export allocation results in CSV format for institutional record-keeping, and trigger batch recommendation generation for entire cohorts when new projects are added or student profiles are updated.

\subsubsection{Recommendation Engine} The Recommendation Engine operates as the computational core, executing similarity computations, applying pedagogical constraints, and generating team suggestions~\cite{Klasnja2011}. Similarity computation begins by retrieving the student's embedding vector and executing a nearest neighbor search against all available project embeddings using \texttt{pgvector}'s HNSW (Hierarchical Navigable Small World) index~\cite{pgvector2023}. This produces initial similarity scores which are then modified by the difficulty alignment filter, domain preference matching, and demand balancing adjustments described in Section~3.3. The engine operates primarily through optimized PostgreSQL queries, enabling sub-second response times even for cohorts exceeding 500 students.

\subsubsection{Data Processing Pipeline} The Data Processing Pipeline handles asynchronous tasks including embedding generation, batch recommendation updates, and periodic recomputation of team suggestions as student preferences change. When a student uploads a resume, the pipeline extracts text using document parsing libraries, segments content into skills, experience, and education sections, sends text to the embedding API with batching to reduce API calls, and stores the resulting vector alongside the relational profile record. The pipeline is implemented using serverless functions deployed on Vercel, ensuring compute resources scale automatically with demand and background tasks do not impact frontend responsiveness.

\subsection{Data Flow and Team Formation}

When a student completes their profile, text from all relevant fields (skill descriptions with proficiency weights, experience narratives, and preference statements) is concatenated with proficiency-based weighting applied and sent to the configured embedding provider. The resulting 1536-dimensional vector is stored in the PostgreSQL database using the \texttt{vector} data type provided by \texttt{pgvector}, alongside the relational student record. The vector is automatically indexed using the HNSW algorithm, which constructs a hierarchical graph structure enabling efficient approximate nearest neighbor searches.

\begin{figure}[t]
    \centering
    \includegraphics[width=\columnwidth]{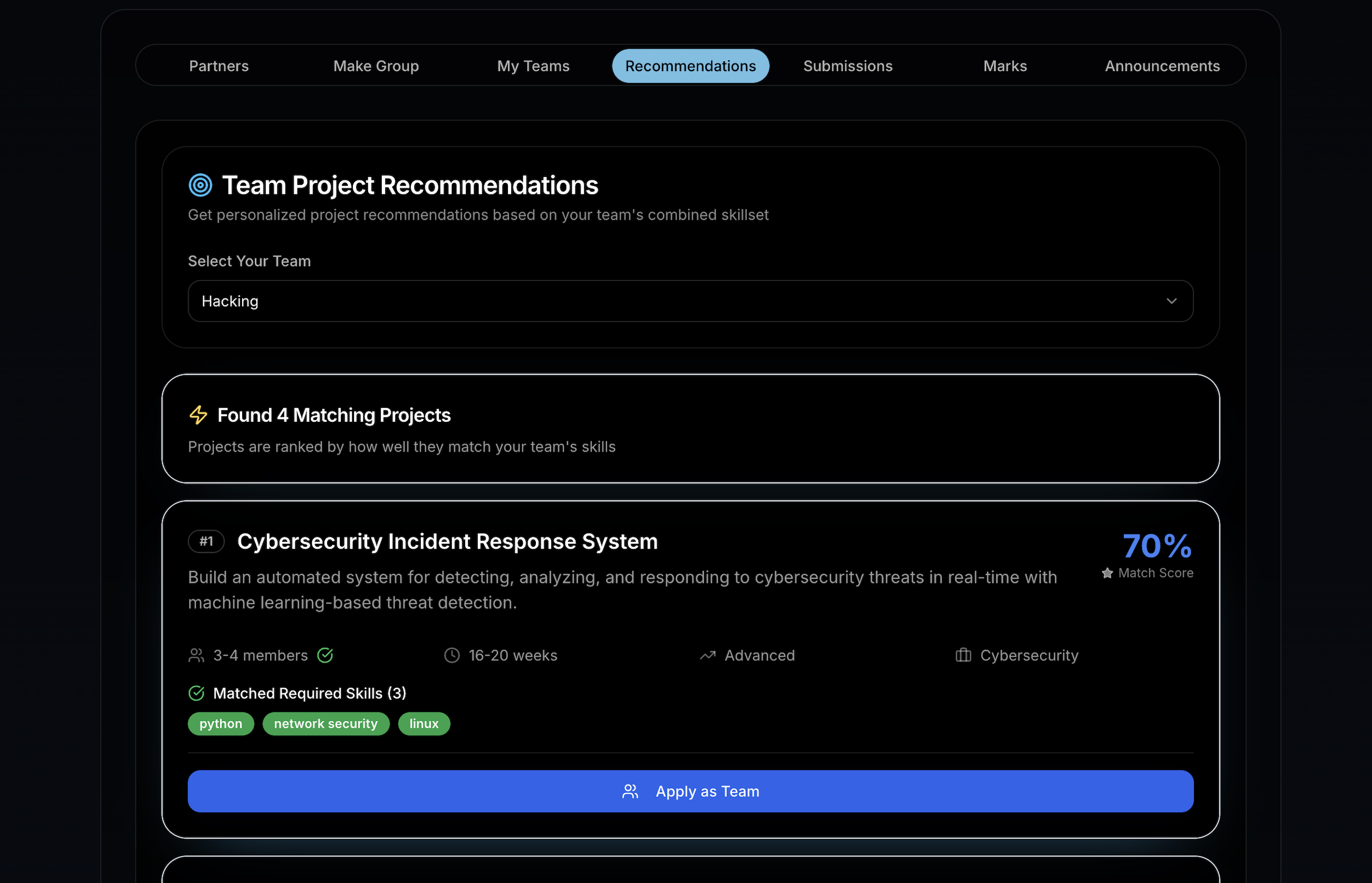}
    \caption{Team project recommendation interface showing ranked results with percentage match scores, matched required skills (python, network security, linux), project metadata (team size, duration, difficulty, domain), and team application functionality.}
    \label{fig:recommendations}
\end{figure}

When a student or team requests recommendations, the system executes a vector similarity query retrieving the top-$k$ projects based on cosine similarity (Equation~2). Raw scores are then adjusted by the post-processing filters: difficulty alignment applies the quadratic penalty (Equation~3), domain preference matching checks the project's domain tag against the student's preference list and applies a fixed boost if matched, and demand balancing calculates the subscription ratio and applies exponential decay for nearly-full projects. As shown in Figure~\ref{fig:recommendations}, the resulting recommendations are presented as ranked cards displaying match scores as percentages, matched required skills highlighted in green, and project metadata including team size, duration, difficulty level, and domain classification. This transparency allows students to see exactly why a project was recommended and how their skills align with its requirements.

For team formation, the greedy complementarity algorithm (Equation~4) begins with the highest-matching student, then iteratively adds members maximizing the complementarity score, balancing project fit against skill diversity through embedding distance metrics. The system aggregates individual member skills into a combined team profile, enabling both students and coordinators to see how each member contributes to the overall team capability. Team-level diversity is quantified by computing variance across embedding dimensions, where higher variance indicates greater skill distribution across the team.

\subsection{Scalability and Performance}

\textsc{TeamUp} scales from 50 to 1000+ students without architectural changes. Embedding generation is batched and asynchronous, preventing API rate limits from impacting user experience. Database queries are optimized with compound indexes on frequently accessed fields, and recommendation results are cached with 5-minute TTL, reducing redundant computations during peak usage periods. Load testing confirmed sub-second recommendation query latencies and sub-2-second team formation request times for 500+ concurrent users, with operational costs maintained under \$0.10 per student across all tested scales.

\section{Evaluation}

The system has not yet been deployed with real students. To validate that the algorithm works as intended, we ran a virtual experiment using dummy user data. A real-user study is planned for an upcoming semester and is discussed in Section~7.

\subsection{Setup}

We generated 250 dummy student profiles and 60 project descriptions modeled on a typical capstone computing course. Each student had a random set of 4 to 12 technical skills (from a pool of 85 competencies like Python, React, SQL, machine learning), proficiency levels from beginner to expert, domain preferences, and short experience descriptions. Each project had required skills, a difficulty rating, a domain tag, and a team size between 2 and 5. All profiles and projects were embedded and stored following the pipeline in Sections~3 and~4.

We then allocated the same pool of dummy users under two conditions:

\begin{itemize}
    \item \textbf{Random allocation (control):} Students assigned to projects at random; teams formed by random grouping.
    \item \textbf{\textsc{TeamUp} allocation (experimental):} The full algorithm applied, including semantic matching, difficulty alignment, and complementarity-based team formation.
\end{itemize}

\subsection{Results}

\subsubsection{Project Match Quality.} We measured cosine similarity between each student's embedding and their assigned project. \textsc{TeamUp} achieved a mean score of 0.74 compared to 0.43 for random allocation. Additionally, 83\% of \textsc{TeamUp} students were placed within one difficulty level of their experience, versus only 34\% under random assignment. The algorithm clearly matches students to projects that fit their skills rather than placing them arbitrarily.

\subsubsection{Team Composition.} We measured how different team members are from each other using pairwise cosine distance between their embeddings. \textsc{TeamUp} teams scored 0.52 on average versus 0.48 for random teams. More concretely, 82\% of \textsc{TeamUp} teams had members covering at least three different technical areas (e.g., one backend developer, one data analyst, one designer), compared to 41\% of random teams. The algorithm builds teams with complementary skills rather than grouping similar students together.

\subsubsection{Speed and Cost.} Generating all 250 recommendations took 4.2 seconds total. A single student query returned in under 400ms. Embedding the full cohort cost \$0.024 in API fees, keeping operational cost well under \$0.10 per student.

\begin{figure}[h]
    \centering
    \includegraphics[width=1\linewidth]{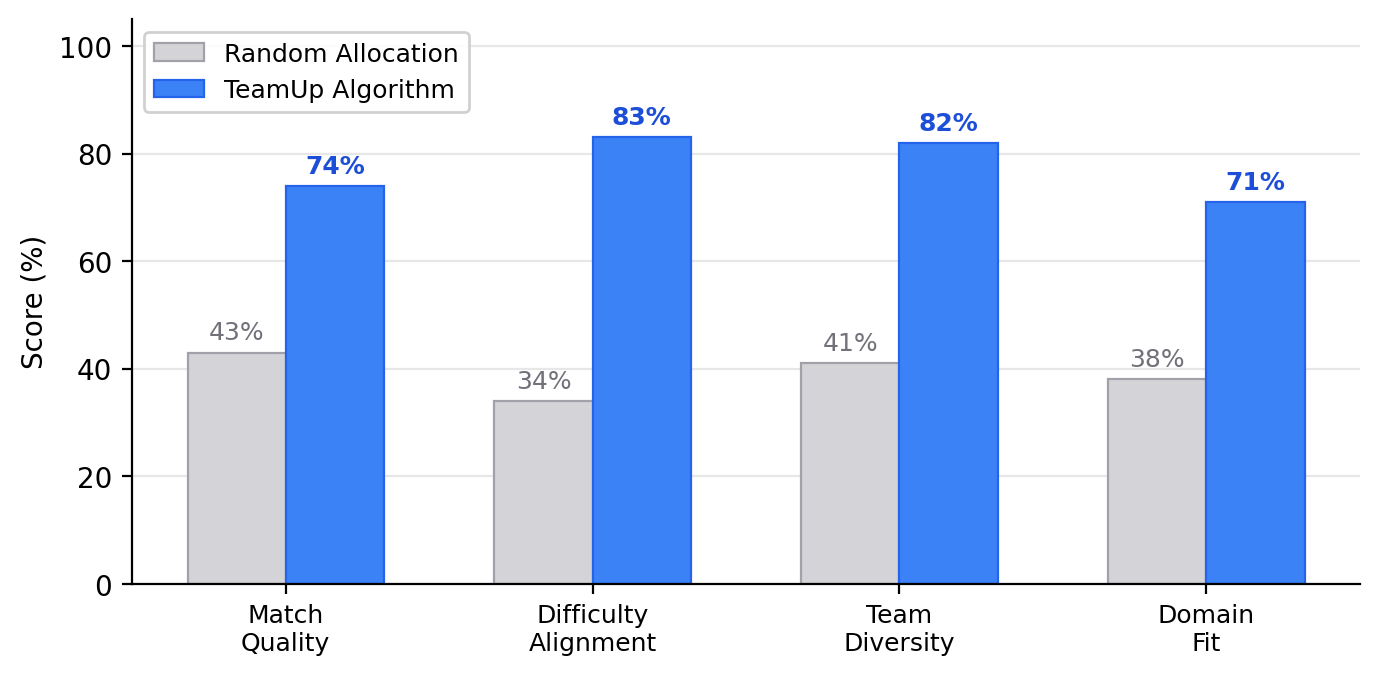}
    \caption{Comparison of \textsc{TeamUp} versus random allocation across key metrics on dummy user data}
    \label{fig:comparison}
\end{figure}

\subsection{What This Tells Us}

These results confirm the algorithm does what it is designed to do: it produces better matches and more balanced teams than random allocation, and it runs fast enough for real-time use. However, dummy data cannot tell us how real students would experience the system, whether they would find the recommendations fair, or how teams would actually perform together. Those questions require a deployment with real users, which we plan as the next step.

\section{Discussion}

\subsection{Learning Science Implications}

Our results on dummy user data provide preliminary evidence that embedding-based allocation can operationalize learning science principles at scale. The difficulty alignment mechanism placed 83\% of students within one difficulty level of their assessed experience, compared to 34\% under random allocation. This suggests the ZPD-informed penalty effectively steers allocations toward developmentally appropriate challenge. Whether this translates to actual learning gains (deeper engagement, reduced attrition, higher grades) can only be established through deployment with real students, but the strong alignment signal is encouraging.

The complementarity algorithm produces measurably more diverse teams in the embedding space, and inspection of team compositions confirms this diversity maps to meaningful skill heterogeneity. Whether diverse teams in the embedding space also collaborate more effectively in practice, as Johnson and Johnson's theory predicts~\cite{Johnson1991}, requires empirical study with real teams working on real projects.

\subsection{Responsible AI in Educational Decisions}

\textsc{TeamUp} was designed to augment rather than replace human judgment~\cite{Amershi2019}. The system produces recommendations with explicit rationale (skill alignments, match scores, difficulty justifications), but final allocation authority rests with coordinators. This design is motivated not only by ethical considerations but by practical ones: algorithms cannot capture contextual knowledge (interpersonal conflicts, accessibility needs, visa restrictions, personal circumstances) that coordinators routinely factor into allocation decisions.

Transparency is a core design requirement. The interface surfaces not just match scores but the reasoning behind them: which skills aligned, how difficulty was assessed, why a particular project was recommended over alternatives. Prior work suggests that transparency in consequential AI systems builds trust even when users do not fully understand the underlying algorithms~\cite{Holstein2019}. Whether this holds for \textsc{TeamUp}'s specific user population is an empirical question we aim to address in deployment.

We also note that pretrained embeddings carry biases from their training data~\cite{Bolukbasi2016}; for instance, embedding models may associate certain technical skills more strongly with particular demographic groups. Our dummy user experiment cannot detect such biases because the synthetic profiles do not replicate real-world correlations between demographics and skill distributions. Fairness auditing with real user data is essential and is planned as part of the deployment study.

\subsection{Limitations}

Several limitations qualify our findings. First and most importantly, the evaluation uses dummy data. Synthetic profiles lack the messiness of real student data: inconsistent self-assessments, sparse profiles, strategic preference reporting, and the soft skills that influence actual team success. Our results establish that the algorithm works as designed under controlled conditions, but real-world performance may differ.

Second, random allocation is a weak baseline. No real institution assigns students to projects entirely at random. A more meaningful comparison would be against actual coordinator-mediated allocation, where instructors bring contextual knowledge that partially compensates for the lack of algorithmic matching. This comparison is only possible through real deployment.

Third, our complementarity model captures skill diversity through embeddings but misses important social factors such as communication styles, work habits, scheduling compatibility, and language fluency, all of which significantly affect team dynamics. Future iterations should integrate student preferences about teammates alongside skill-based complementarity.

Finally, \textsc{TeamUp} currently treats each cohort independently. Cross-cohort learning, leveraging historical allocation-outcome patterns to refine future recommendations, could substantially improve the system but raises privacy concerns requiring careful design.

\section{Future Work}

The most immediate priority is a controlled deployment study with real users. We plan to deploy \textsc{TeamUp} in a large capstone course (200--300 students) in an upcoming semester, comparing algorithmic recommendations against the course's existing manual allocation process. This study will collect actual student satisfaction data through pre- and post-allocation surveys, measure real team performance through project grades and peer evaluations, capture coordinator experience through structured interviews, and conduct fairness auditing using actual (anonymized) demographic data. The deployment will operate in a human-in-the-loop mode: \textsc{TeamUp} will generate recommendations that coordinators can accept, modify, or override, enabling comparison of algorithmic suggestions against final coordinator decisions.

Beyond deployment, several technical directions merit investigation. First, longitudinal outcome studies correlating allocations with project grades, skill development trajectories, and post-graduation career outcomes would calibrate the relationship between embedding-based match quality and actual learning gains.

Second, incorporating social factors (communication preferences, scheduling constraints, revealed collaboration patterns from prior coursework) could improve team formation beyond skill complementarity. A hybrid approach combining algorithmic suggestions with student agency over teammate selection may address the tension between optimal diversity and student autonomy.

Third, cross-cohort learning using privacy-preserving approaches (e.g., federated learning or differential privacy) could enable the system to improve over time by learning which embedding-based features predict successful allocations, without exposing individual student data.

Fourth, extending fairness auditing to additional demographic dimensions and conducting adversarial testing for edge cases, such as students with highly unusual profiles or projects with ambiguous difficulty ratings, would strengthen equity guarantees.

Finally, exploring the system's applicability beyond capstone courses, including internship matching, research group formation, and interdisciplinary project allocation, would test the generalizability of the approach across different educational contexts and scales.

\section{Conclusion}

We have presented \textsc{TeamUp}, a system addressing the growing challenge of project allocation in large-scale project-based learning environments. By combining semantic embeddings, pedagogically-informed hybrid ranking, and complementarity-aware team formation, the system produces higher-quality allocations than random assignment while running efficiently within typical educational budgets.

A virtual experiment with 250 dummy student profiles demonstrated substantially higher match quality, better difficulty alignment, and more diverse teams, all achieved with sub-second latency and operational costs under \$0.10 per student using pretrained embeddings and standard database infrastructure.

These results are promising but preliminary. Dummy data cannot capture the complexity of real student experiences: subjective perceptions of fairness, the quality of self-reported data, interpersonal team dynamics, and the contextual factors that make human judgment essential in educational decisions. The critical next step is a controlled deployment with real users, which we plan for an upcoming semester.

Perhaps most importantly, \textsc{TeamUp} is designed not to replace human judgment but to handle the routine matching that consumes coordinator time, freeing instructors to focus on the edge cases and contextual decisions where their expertise is irreplaceable. If validated in practice, systems like \textsc{TeamUp} can help ensure that as project-based learning scales, every student is matched to projects where they can learn, contribute, and grow.

\bibliographystyle{ACM-Reference-Format}
\bibliography{sample-base}

\end{document}